\documentclass[12pt]{article}
\usepackage{epsfig}

\begin{document}

\begin{center}
{\Large Multi-gluon field approach of QCD \footnote{Presented at the
XXX. Mazurian Lakes Conference, Sept.~2-9, 2007, Piaski, Poland}}
\vspace{0.3cm}

H.P. Morsch\\
Institut f\"ur Kernphysik, Forschungszentrum J\"ulich, D-52425
J\"ulich, Germany\\ and Soltan Institute for Nuclear Studies, Pl-00681
Warsaw, Poland \\P. Zupranski \\
Soltan Institute for Nuclear Studies, Pl-00681 Warsaw, Poland
\end{center}

\begin{abstract}
The decay of 2-gluon colour singlets in quarks: $2g\rightarrow q\bar q +2q2\bar
q$ has been simulated with the Monte-Carlo method, taking into account 
an effective 1-gluon exchange interaction between the emitted quarks, which was
folded with a 2-gluon density determined self-consistently. 2-gluon
densities were found with different radii, which correspond to $0^{++}$
glueballs of the size of light $q\bar q$, $s\bar s$, $c\bar c$, $b\bar
b$ and heavier $q\bar q$ systems.   
Binding potentials between the two gluons have been deduced, which
are consistent with the confinement potential from lattice results. 
However, self-consistency for the deduction of 2-gluon
densities requires {\bf massless} (or very light) quarks for all
flavours. The masses are given by the binding energies of quarks and gluons,
yielding excitation spectra of $0^{++}$ glueballs and $\Phi$,
$J/\Psi$ and $\Upsilon$ states consistent with observation. 
The sum of q-q potentials yields a strong coupling $\alpha_s$
consistent with the available data up to large momenta. \\
The nucleon is described by a gluonium state coupled
to 3 valence quarks, yielding ground state and radial excitations
consistent with experiment. Finally, we discuss the compressibility of
the nucleon and relate it to that of nuclear matter.  
\end{abstract}
  
\begin{center}
{\large \bf 1. Introduction}
\end{center}

The two key problems in the understanding of the strong interaction
are the confinement of quarks and gluons and the origin of mass, both
related to the non-perturbative structure of quantum chromodynamics
(QCD). A linearly rising confinement potential between quarks has been
derived in potential models~\cite{spec,relqm} and lattice QCD
simulations~\cite{Bali,BBali}, but its origin is not well
understood. The mass term in the QCD Lagrangian is also not
understood, but for the generation of mass a coupling of the quarks to
a scalar Higgs background field has been proposed. Finite quark
masses give rise to the axion problem, which has not been solved.

For the description of QCD in the non-perturbative regime mainly two
non-perturbative methods have been applied, solutions of
Dyson-Schwinger equations~\cite{DS} and lattice QCD~\cite{latt}, which
solves the QCD equations by path integral methods on a space-time
lattice. 

\begin{center}
{\large \bf 2. Deduction of 2-gluon densities}
\end{center}

In this paper a new phenomenological method is presented,
which starts from the conjecture, that the non-Abelian structure of
QCD may generate bound 2-gluon systems, which  decay into $q\bar q$ pairs.  
For the description of such bound states $\Phi$ we  write the
radial wave functions in the form 
$\psi_{\Phi}(\vec r=\vec r_1-\vec r_2)=[\psi_1(\vec r_1)\ \psi_2(\vec
r_2)]$,
where $\psi_j(\vec r_j)$ are the radial wave functions of the two gluons.
To investigate the properties of such 2-gluon systems
we studied the decay $2g\rightarrow q\bar q + 2q2\bar q$ with an
attractive interaction between the emitted quarks. 

Assuming an effective 1-gluon exchange interaction $V_{1g}(R)=-\alpha_s/R$ between the
emitted quarks with relative distance $R=|\vec r_i-\vec r_j|$,
the decay from a 2-gluon system $2g\rightarrow (q\bar q)^n$ requires a
modification of the free q-q interaction by the density of the 2-gluon
system, which may be expressed by a folding integral 
\begin{equation}
\label{eq:qq}
V_{qq}(R)=\int d\vec r\ \rho_{\Phi}(\vec r\ )\ V_{1g}(\vec R-\vec r\ )\ ,
\end{equation}
where $\rho_{\Phi}(\vec r\ )$ is the 2-gluon density $\rho_{\Phi}(\vec r\
)=|\psi_{\Phi}(\vec r\ )|^2$. 

It is interesting to note, that for a spherical density the Fourier transform
of eq.~(\ref{eq:qq}) to momentum (Q) space yields 
\begin{equation}
\label{eq:qqQ}
V_{qq}(Q)=-\frac{4\pi \alpha_s}{Q^2}\ \rho_{\Phi}(Q)\ ,
\end{equation}
where $\rho_{\Phi}(Q)=4 \pi \int r^2 dr\ j_o(Qr)\ \rho_{\Phi}(r)$.
Comparing this with the standard 1-gluon exchange force yields a Q-dependent
strong coupling $\alpha_s(Q)= \alpha_s\ \rho_{\Phi}(Q)$,
which is qualitatively consistent with 
the known fact of a ``running'' of $\alpha_s(Q)$ and the condition 
$\alpha_s(Q)$$\rightarrow$ 0 for $Q$$\rightarrow$$ \infty$ (asymptotic
freedom). 

Further, finite size of the decaying 2-gluon system have been taken
into account, as well as the fact, that the decay into $2q 2\bar q$
favours relative angular momentum L=0 between the emitted quarks,
whereas for the decay in $q \bar q$ the outgoing quarks are in a
relative p-state (L=1). Details are given in ref.~\cite{MZ}.
By relativistic Fourier transformation~\cite{Kelly} the effective
interaction~(\ref{eq:qq}) can be transformed to momentum space 
\begin{equation}
\label{eq:qq2}
V_{qq}(Q')= 4\pi \int R^2 dR\ j_o(Q' R)\ V_{qq}(R)\ ,
\end{equation}
with $Q'=Q\ \sqrt{1+[Q^2/4m_{\Phi}^2]}$ and
$m_{\Phi}$ being the mass of the 2-gluon system. 

\begin{figure} [ht]
\centering
\includegraphics [height=12.8cm,angle=0] {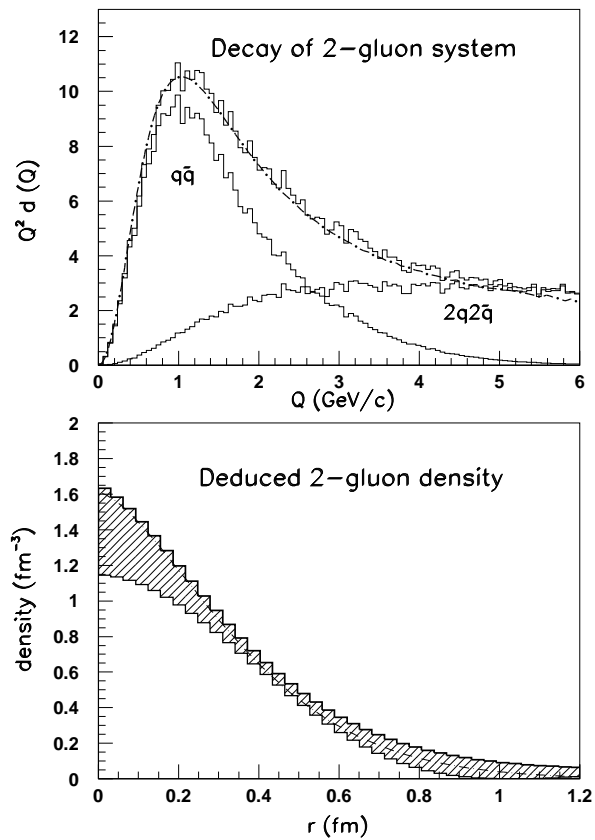}
\label{fig1}
\caption{Upper part: Resulting 2-gluon momentum distributions (multiplied by
  Q$^2$) for decay in $q\bar q$ and $2q2\bar q$ and sum.
Lower part: Deduced 2-gluon density with estimated error band.} 
\end{figure}

Monte-Carlo simulations of gluon-gluon scattering have been performed
in fully relativistic kinematics, in which the 2 gluons in the final state  
can decay in $q \bar q$ and $2q 2\bar q$ (using massless quarks).
The potential  $V_{qq}(\Delta p)$~(\ref{eq:qq2})  
has been used as a weight function between the outgoing quarks (with the
relative momenta $\Delta \vec p=\vec p_i - \vec p_j$). Resulting gluon
momentum distributions $d_{q\bar q}(Q)$ and $d_{2q2\bar q}(Q)$ for
decay into  $q \bar q$ and $2q 2\bar q$ were generated. Their sum
$D_{\Phi}(Q')=d_{q\bar q}(Q') +d_{2q2\bar q}(Q')$ can be
related to the radial density $\rho_{\Phi}(r)$ of the 2-gluon system  
\begin{equation}
\label{eq:distr}
D_{\Phi}(Q')=4 \pi \int r^2 dr\ j_o(Q'r)\ \rho_{\Phi}(r)\ ,
\end{equation}
with $Q'$ as in eq.~(\ref{eq:qq2}). 

The condition, that $\rho_{\Phi}(r)$ in the interaction~(\ref{eq:qq2})
and in eq.~(\ref{eq:distr}) should be the same, allowed us to determine
this density. Resulting momentum distributions $d_{q\bar q}(Q)$ and $d_{2q2\bar
 q}(Q)$ for a self-consistent solution with $<r^2>\approx$0.5 fm$^2$
are given in the upper part of fig.~1. We see that the sum
$D_{\Phi}(Q)$ is in reasonable agreement with 
$\rho_{\Phi}(Q)$ from the Fourier transformation of $\rho_{\Phi}(r)$ inserted
in eq.~(\ref{eq:qq2}) (dot-dashed line), which is quite well
approximated by a radial dependence $\psi_{\Phi}(r)=\psi_o\
exp[-(r/a)^{\kappa}]$ with values of $\kappa$ of about 1.5. The
resulting density is given in the lower part of fig.~1, which
indicates clearly that a self-stabilized 2-gluon field is generated.
The mass $m_{\Phi}$ in the relation between $Q$ and $Q'$ has been used
as a fit parameter; for a 2-gluon system with a mean square radius of
about 0.5 fm$^2$ this yields $m_{\Phi}\sim$ 0.68 GeV . This is 
consistent with the gluon pole mass of $0.64\pm 0.14$ GeV deduced in
ref.~\cite{Alex}. We shall see, that the extracted mass can be understood as 
binding energy of the 2-gluon system including relativistic mass corrections.
 
The extracted 2-gluon density should give a significant contribution to the
gluon 2-point functions extracted from lattice QCD simulations in form of
gluon field correlators~\cite{FC,DiG} and the QCD gluon propagator
(see e.g.~\cite{Mand}). In the upper part of fig.~2 we make a
comparison of our results with the 2-gluon field correlator $C_{\perp}(r)$  
of Di~Giacomo et al.~\cite{DiG}, in the lower part with the gluon
propagator of Bowman et al.~\cite{Bow04}. 

We get a good agreement with the lattice data (note that in fig.~2 the gluon
propagator is multiplied by $Q^2$), but we have to add a second
2-gluon component of smaller size, given by the dot-dashed lines. 

\begin{center}
{\large \bf 3. Bindung potential of gluons and $0^{++}$ glueball states}
\end{center}

\begin{figure} [pt]
\centering
\includegraphics [height=13.2cm,angle=0] {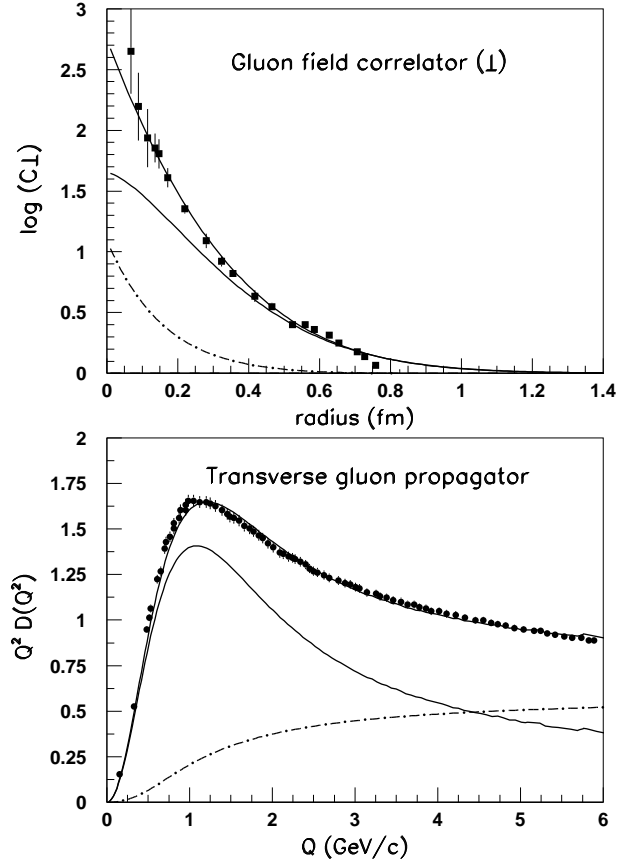}
\label{fig2}
\caption{Gluon field correlator $log(C_{\perp})$ from ref.~\cite{DiG} (upper
  part) and gluon propagator from ref.~\cite{Bow04} (lower part)
  from lattice QCD simulations in 
  comparison with our results. The lower solid lines correspond to the
  density $\rho_{\Phi}(r)$ and its Fourier transform $\rho_{\Phi}(Q)$,
  respectively, and the dot-dashed lines to an additional vector component.
 The sum of both contributions yields a good description of both lattice data.} 
\end{figure}
\begin{figure} [pt]
\centering
\includegraphics [height=13.7cm,angle=0] {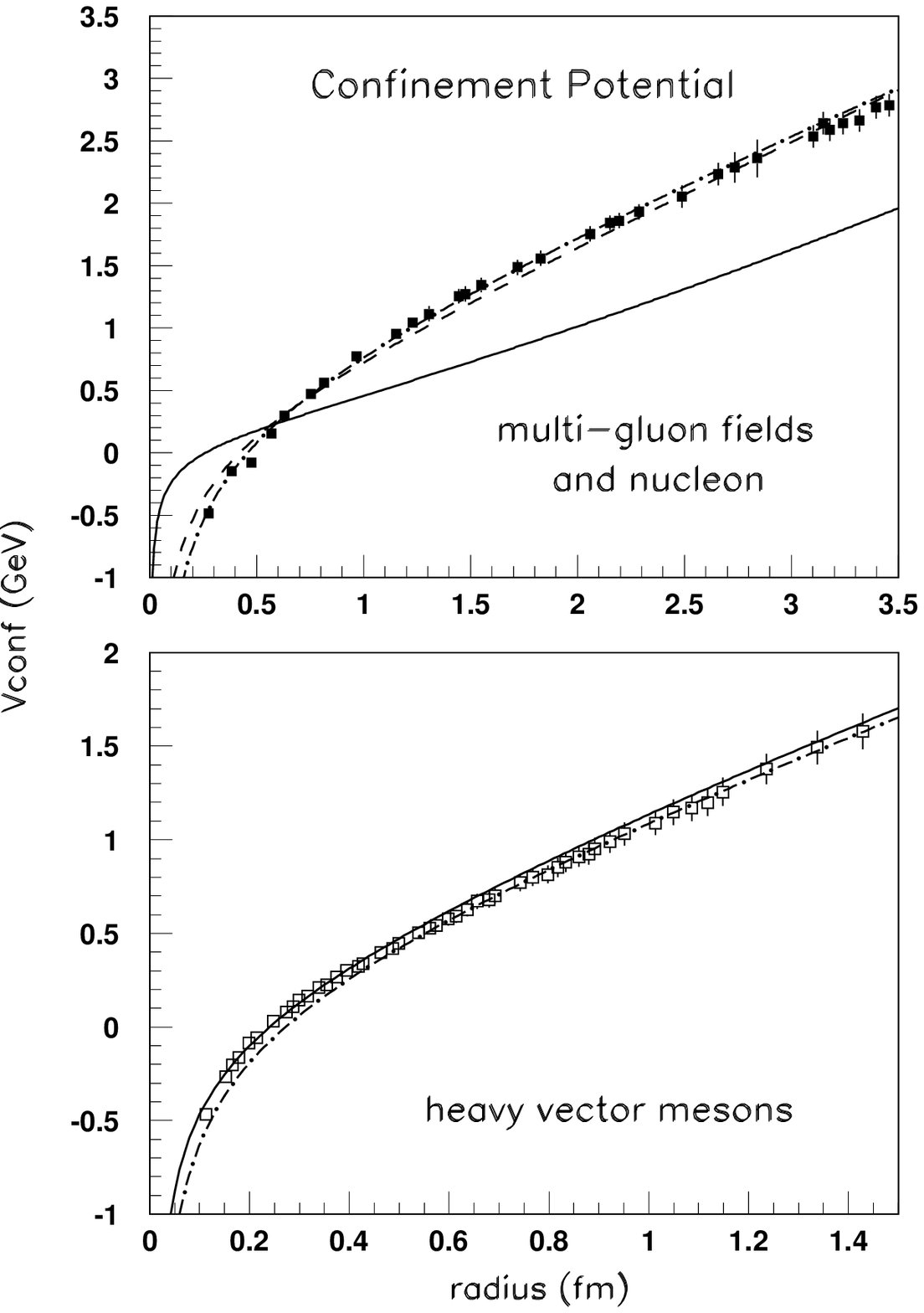}
\label{fig3}
\caption{Upper part: deduced 2-gluon binding potentials (dot-dashed
  and dashed lines) in comparison with the confinement potential from lattice gauge
  calculations~\cite{BBali} (upper part). The solid line corresponds to
  the binding potential for the nucleon discussed below.
  Lower part: potentials deduced for heavy vector ($s\bar s$, $c\bar c$, and
  $b\bar b$) systems discussed below in comparison with lattice QCD
  results~\cite{Bali}.} 
\end{figure} 
A finite 2-gluon density as shown in fig.~1 may be interpreted as a
bound state of the two gluons (glueball). Therefore, from the  
2-gluon density the binding potential of the 2-gluon system can
be obtained by solving a three-dimensional reduction of the 
Bethe-Salpeter equation in form of a relativistic Schr\"odinger equation 
\begin{equation}
-\bigg( \frac{\hbar^2}{2\mu_{\Phi}}\ \Big [\frac{d^2}{dr^2} +
  \frac{2}{r}\frac{d}{dr}\Big ] - V_{\Phi}(r)\bigg) \psi_{\Phi}(r) =
  E_i \psi_{\Phi}(r)\ ,  
\label{eq:4}
\end{equation}
where $\psi_{\Phi}(r)$ is the 2-gluon wave function and
$\mu_{\Phi}$ a relativistic mass parameter, which is related to
$m_{\Phi}$ by $\mu_{\Phi}=\frac{1}{4}\ 
m_{\Phi}+\delta m$, where $\delta m$ is a relativistic correction. 
Slightly different solutions of the binding potential were obtained, which
are given by the dot-dashed and dashed lines in the upper part of fig.~3. 
Because of the relation $2g\rightarrow (q\bar q)^n$ this potential can also be
considered as confinement potential between the emitted quarks. This
is in surprising agreement with the 1/r + linear form expected from
potential models~\cite{spec,relqm} and consistent with the confinement
potential from lattice QCD~\cite{BBali}. It is important to note, 
that our potential reproduces the  1/r + linear
form without any assumption on its distance behavior; this is entirely
a consequence of the deduced radial form of the 2-gluon wave function.
 
Bound state energies $E_i$=0.68$\pm$0.10 GeV, 1.70$\pm$0.15 GeV, and
2.58$\pm$0.20 GeV have been extracted. Further, in the q-q
potential~(\ref{eq:qq2}) we find one bound state with an energy in the
order of -10 MeV. This very low binding energy indicates clearly, that
the glueball states must 
have a large width, since we expect $\Gamma\sim 1/E_o$. This is
consistent with the general expectation for the width of glueball
states ($\Gamma\ge$ 500 MeV). 
From these results we may conclude, that the glueball ground state with 
$E_o$=0.68$\pm$0.10 GeV and a large width may be identified with the scalar
$\sigma$(600).  

Glueball masses have been deduced also from lattice
simulations~\cite{Bali,Morning}, in which a glueball mass below 
1 GeV has not been found. However, in these simulations $0^{++}$ glueball
masses have been extracted at about 1.7 and 2.6 GeV, which correspond
very nicely to the first and second radial excitation in table~1. Our
evidence for a low lying glueball is supported by QCD sum rule
estimates~\cite{Narison}, which also require the existence of a low lying gluonium
state below 1 GeV.

\begin{center}
{\large \bf 2. Heavy flavour neutral systems and $\alpha_s(Q)$}
\end{center}

Self-consistent 2-gluon densities have been deduced also for smaller
systems corresponding to the size of $s\bar s$, $c\bar c$, and $b\bar b$
mesons. Resulting momentum distributions and the corresponding Fourier
transformed densities are given in fig.~4, which are in good agreement.
\begin{figure} [ht]
\centering
\includegraphics [height=12.8cm,angle=0] {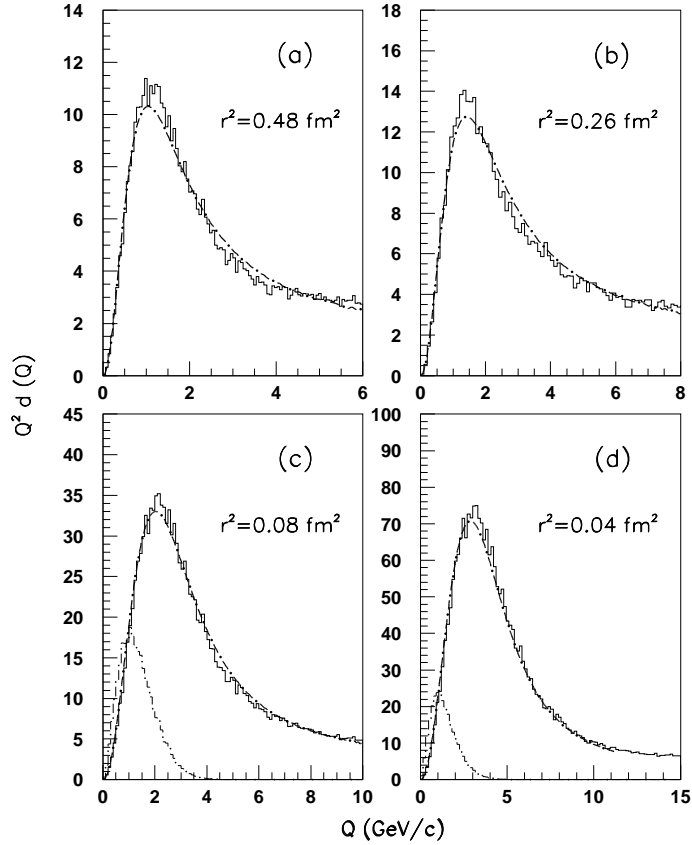}
\label{fig4}
\caption{Momentum distributions (histograms) and Fourier transforms of the 2-gluon
  density (dot-dashed lines) for 4 different densities investigated, corresponding to
  light $q\bar q$ system (a), $s\bar s$ (b), $c\bar c$ (c), and
  $b\bar b$ (d). The dot-dashed histograms are simulations assuming 
  c and b quark masses of 1.4 and 4.5 GeV, respectively.}
\end{figure}

It is interesting to investigate the effect of quark masses in our
simulations. Using quark masses of 1.24, and 4.5 GeV for c and b
quarks, respectively, yields the lower dot-dashed histograms,
indicating that self-consistent solutions are not
possible. Thus, for all systems the intrinsic quark masses have to be
zero (or very small). The 2-gluon binding potential is given in the
lower part of fig.~3 in good agreement with the confinement potential
from lattice QCD~\cite{Bali}. 

Since all intrinsic quark masses have to be small in our approach, the
masses of the different systems have to be explained in a different
way. Whereas the binding potential (of 2-gluons) gives rise to binding
energies in the order of 1 GeV, the binding potential of 
quarks~(\ref{eq:qq}) depends strongly on the size of the 2-gluon
densities. Therefore, the binding of quarks can be much larger. This
is shown for the different systems in fig.~5. 
\begin{figure} [ht]
\centering
\includegraphics [height=12.5cm,angle=0] {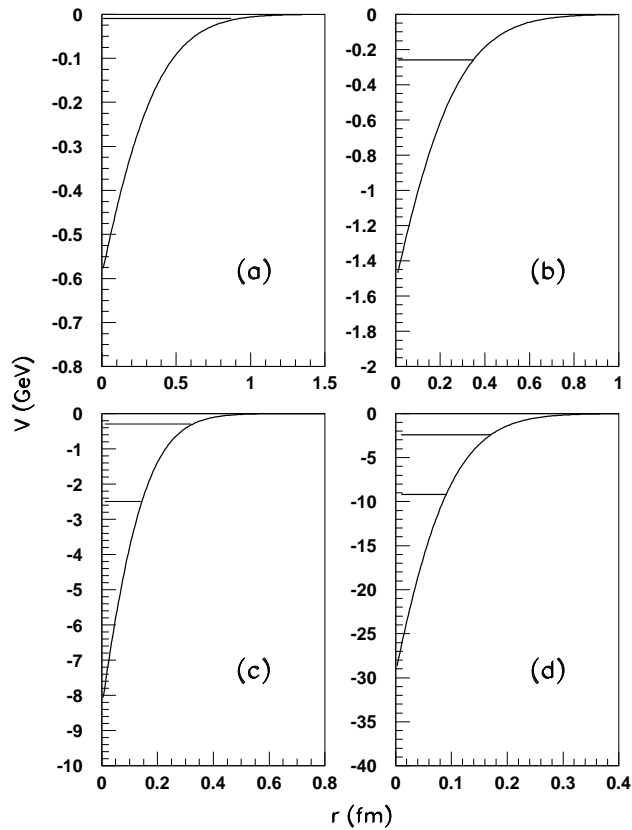}
\label{fig5}
\caption{Folding potential~(\ref{eq:qq}) for the four different
  cases (a) -- (d) corresponding to light glueballs,
  $s\bar s$, $c\bar c$, and $b\bar b$ with the binding energies indicated.}
\end{figure}
For the heavy systems (which are of small size) the binding energies
are in the order of 2.4 and 9.0 GeV, respectively, which shows that indeed the
masses of all systems can be explained by the binding of quarks and gluons.
The resulting energies of ground and radial $\Phi$, $\Psi$ and
$\Upsilon$ states are in good agreement with the experimental
spectra. Details of these calculations will be given elsewhere.

The Fourier transformed potential~(\ref{eq:qqQ}) is directly related to the
strong coupling $\alpha_s(Q)$. Using the different 2-gluon density
distributions $\rho_i(Q)$ deduced from fig.~4 we obtain
$\alpha_s=\Sigma_i\ a_i \rho_{\Phi_i}(Q)$. This gives a 
quantitative description of $\alpha_s$ up to large momenta, see fig.~6.
\begin{figure} [ht]
\centering
\includegraphics [height=9.5cm,angle=0] {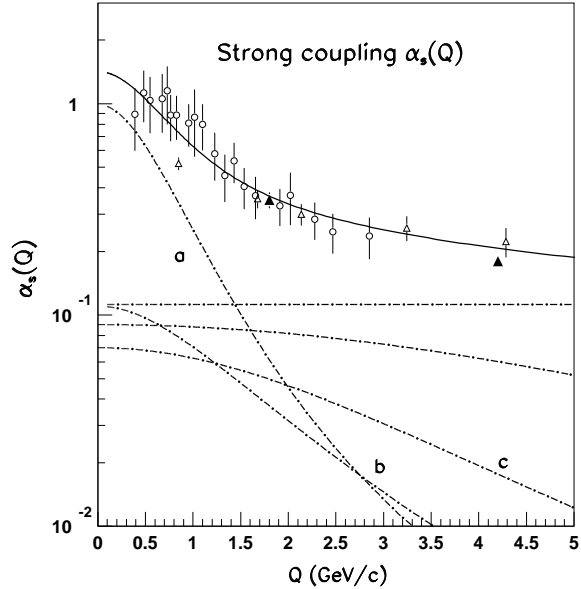}
\label{fig6}
\caption{Strong coupling $\alpha_s$ from lattice QCD~\cite{Fu} and
  experiment~\cite{LQCD} (triangles) in comparison with our results, given by the solid
  line. The contributions of the different 2-gluon densities
  $\rho_{\Phi_i}(Q)$ are given also.} 
\end{figure}

\begin{center}
{\large \bf 5. Nucleon structure}
\end{center}

Baryons may be described in our approach assuming the decay
$4g\rightarrow 5(q\bar q)\rightarrow (3q\ q\bar q) +(3\bar q\ q\bar
q)$, which means 4g $\rightarrow$ (baryon + antibaryon). 
Thus, we decribe the nucleon by 3 valence quarks coupled to a 2-gluon
field yielding $\rho_N(r)=4\pi \int \rho_{3q}(r')\rho_{\Phi}(r-r')dr'$. The
resulting binding potential is given in the upper part of fig.~3 by
the solid line, which is more shallow than the confinement potential,
but the attraction between the emerging quarks is increased by a
factor 9. The binding energies of the nucleon g.s.~and radial
excitations are 0.94 GeV, 1.42$\pm$0.07 GeV, and 1.82$\pm$0.12 GeV in
good agreement with experiment. 

\begin{center}
{\large 5.1. Compressibility: from nucleon to nuclear matter}
\end{center}

Finally, we discuss the nucleon compressibility, which may be linked to
that of nuclear matter. 
From the excitation of the first~radial state, the ``breathing mode'' of
the nucleon, the compressibility $K_N$ has been extracted by operator sum
rules~\cite{Morsch,Mo} yielding values of about 1.3 GeV. The breathing
mode has also been investigated in high energy p-p and $\pi$-p
scattering~\cite{MoZuneu}. From a comparison of transition densities
deduced from inelastic p-p and e-p scattering strong multi-gluon
contributions were extracted, which were about a factor 4 stronger
than those of the valence quarks. This is in good agreement with the
present results. From the multi-gluon potentials deduced in
ref.~\cite{MoZuneu} we may derive the nucleon compressibility
directly. Calculating a potential density for the nucleon given by 
$V\rho_N(r)=\frac{1}{2}\ V_{NN}\ \int \rho_N(r_N)t_{NN}(r-r_N) dr_N$
and adding a kinetic energy term $T$ we obtain the energy density 
$E\rho_N(r)=-(V+T)\cdot \rho_N(r)$. The compressibility is then given
by
\begin{equation}
\label{eq:comp}
K_N=r^2 \frac{d^2 E\rho_N(r)}{dr^2} |_{r=r_o}\ .
\end{equation}
From the analysis of high energy p-p scattering~\cite{MoZuneu} the multi-gluon
potential is well determined.
\begin{figure} [ht]
\centering
\includegraphics [height=9cm,angle=0] {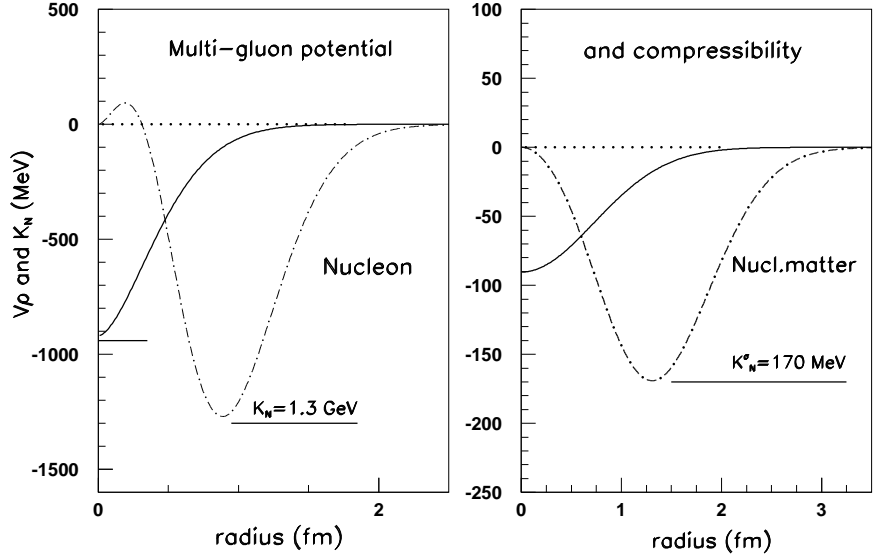}
\label{fig7}
\caption{Left: Nucleon potential density (solid line) and derived
  compressibility function (\ref{eq:comp}) (dot-dashed line).
Right: The same for the scalar nucleon potential, important for
nuclear matter.}
\end{figure}
So, we can determine the compressibility. This is given on the left
side of fig.~7. Indeed, we obtain a compressibility of about 1.3 GeV
consistent with the value obtained from sum rules~\cite{Mo}.

For the case of nuclear matter we assume that the dominant
contribution is due to compressibility of the nucleons (the
compressibility due to the binding of nucleons should
not be much larger than their binding energy). Then the
compressibility is related to the central (scalar) nucleon potential,
which has a mean square radius $\geq$1.5 fm$^2$. Using the
corresponding density the derived compressibility is in the order of
160-170 MeV, this is shown on the right side of fig.~7. This value is
rather close to the compressibility of nuclear matter $K_{\infty}$ of
about 220-250 MeV deduced from the study of the giant monopole
resonance in heavy nuclei. 

\begin{center}
{\large \bf 6. Conclusion}
\end{center}

The present solution of the confinement problem based on our
phenomenological description of two-gluon fields differs entirely 
from earlier suggestions, that confinement could arise from complicated
non-perturbative field configurations (magnetic monopoles, flux tubes,
vortices or strings) in the Abelian projection of QCD, which had severe
problems, e.g.~with Casimir scaling. Further, none of these models could
explain the generation of mass and the complex Yang-Mills gluon
structure found in lattice QCD calculations. In our description all
these problems are tied together and are well described. The masses of
hadrons are explained by binding effects, and all intrinsic quark
masses have to be consistent with zero. Thus, a scalar Higgs field in which the
quark masses are generated is not needed. Also the axion problem does not exist when
quark masses are zero. 

It is of large interest that in our approach, in which many of the
properties of hadrons are well described, including the problem of the
light pion mass and the non-existence of chiral symmetry, quarks arise only from the
decay of multi-gluon systems. This has important consequences for our
understanding of the origin of our universe and of baryogenesis. 

Finally, by the discussion of the compressibility a first example is
given, which shows that the properties of hadrons are strongly tied to
those of nuclear systems.

\end{document}